\documentclass[aps,pre,reprint,superscriptaddress]{revtex4-2}
\usepackage{amsmath,amsfonts,mathtools}

\DeclareMathAlphabet{\mathcal}{OMS}{cmsy}{m}{n}

\usepackage{graphicx}
\usepackage{dcolumn}
\usepackage{bm}
\usepackage{xcolor}
\usepackage{mathptmx}
\usepackage{physics}

\makeatletter
\newcommand{\vast}{\bBigg@{4}}
\newcommand{\Vast}{\bBigg@{5}}
\makeatother

\usepackage{lipsum}

\begin{document}


\title{Survival probability of particles inside the Lemon Billiard}


\author{Daniel Borin}
\affiliation{ \href{https://ror.org/04a5szx83}{University of North Dakota}, 
School of Electrical Engineering and Computer Science, 58202, Grand Forks, ND, USA }
\affiliation{\href{https://ror.org/00987cb86}{São Paulo State University (UNESP)} , IGCE - Physics Department, 13506-900, Rio Claro, SP, Brazil}

\author{Edson Denis Leonel}
\affiliation{\href{https://ror.org/00987cb86}{São Paulo State University (UNESP)} , IGCE - Physics Department, 13506-900, Rio Claro, SP, Brazil}

\author{Diego Fregolent Mendes de Oliveira}
\affiliation{ \href{https://ror.org/04a5szx83}{University of North Dakota}, School of Electrical Engineering and Computer Science, 58202, Grand Forks, ND, USA }



\begin{abstract}

We study the escape of particles in the lemon billiard, a two-parameter family of billiard systems defined by the intersection of two identical circles. 
Using numerical simulations, we explore how the survival probability depends on the position and size of the hole, as well as on the billiard shape parameter. We find that the survival probability exhibits a two-stage decay pattern: an initial exponential regime followed by a long-time power-law tail, a signature of the stickiness effect.
Our results show that the short-time exponential decay rate follows a power-law dependence on the hole size, with different scaling exponents for holes placed in chaotic regions versus mixed phase space regions. For holes located in mixed phase space regions, the decay exponent of the long-time power-law tail remains approximately constant, while the amplitude follows a power-law scaling with hole size. We also examine the dependence of short-time exponential decay rate on the billiard shape parameter and observe scaling behavior for small values of this parameter, which breaks down as the parameter increases.

\end{abstract}


\maketitle

\section{Introduction}

Billiard systems are a type of dynamical system that originally emerged from statistical mechanics. In these systems, a particle moves freely along straight paths (or under the influence of potential forces) within a confined region on a plane \cite{chernov2006chaotic}. When the particle hits the boundary of the billiard, it reflects elastically, changing its velocity according to the reflection law. Essentially, billiards serve as idealized models for situations where particles or waves are restricted to cavities or other uniform regions.

With advances in scientific computation, billiard systems have attracted considerable attention due to their intuitive nature and broad applicability. They provide useful frameworks for addressing complex problems across various scientific disciplines. Applications include biology \cite{boucher2025buzz, PhysRevE.110.054201}, celestial mechanics \cite{de2024analytical, Barutello_2023}, applied mathematics \cite{PhysRevLett.94.100201, PhysRevLett.132.157101}, plasma physics \cite{PhysRevLett.108.064102}, waveguides \cite{PhysRevLett.68.2867} , microwave billiards \cite{PhysRevLett.85.2478}, dispersal of microorganisms in porous media \cite{PhysRevLett.134.188303} and superconductivity experiments \cite{Jonathan_P_Bird_1999}. 

The dynamics of billiard systems depend strongly on the billiard boundary’s geometry, ranging from completely integrable (regular) to chaotic behaviors. A classic example is the circular billiard, whose dynamics are integrable because energy and angular momentum are conserved \cite{chernov2006chaotic}. In this case, the phase space consists of periodic orbits (represented by straight lines). On the other hand, the Bunimovich stadium billiard is characterized by chaotic dynamics. Bunimovich originally studied a stadium billiard formed by two semicircles of radius $R$ connected by two straight segments of length $2a>0$. When $a=0$, the system reduces to the circular billiard \cite{bunimovich1974ergodic}. More recently \cite{PhysRevE.103.012204, PhysRevE.75.046204}, systems with $a<0$ have been studied, where straight segments disappear and the boundary forms two circular sectors resembling a lens or lemon shape. This configuration leads to a mixed phase space, characterized by Kolmogorov–Arnold–Moser (KAM) islands embedded within a chaotic sea.

In this paper, we investigate the survival probability of the lemon billiard with holes of varying sizes placed at different positions on the billiard's boundary. Our focus is on understanding whether this observable exhibits scaling invariance when varying key parameters such as the hole size $h$ and the billiard shape parameter $B$. We analyze the behavior of both the exponential decay at short times and the power-law tails at long times, revealing distinct dynamics depending on the hole’s location in chaotic or mixed regions of phase space.

This paper is organized as follows. In Section~\ref{sec:model}, we introduce the lemon billiard model and discuss the equations governing the system’s dynamics. Section~\ref{sec:survivalprobability} investigates the survival probability of particles escaping through a hole at two different positions, highlighting scaling behaviors under variations of the system parameters. Finally, Section~\ref{sec:conclusion} presents our conclusions.

\section{Lemon Billiard} \label{sec:model}

\begin{figure}[!htb]
	\includegraphics[width=\linewidth]{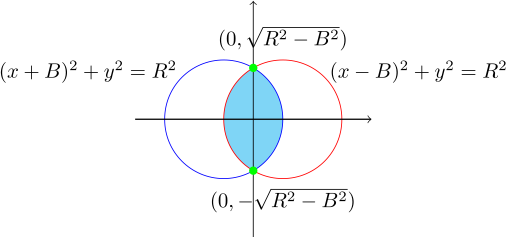}
	\caption{(Color online): Geometry of the lemon billiard. The boundary is formed by the intersection of two identical circles of radius $R$, whose centers are shifted symmetrically by a distance $2B$. The blue and red arcs represent the left- and right-shifted circles, respectively. The cyan area corresponds to the billiard domain. The green dots indicate the points of intersection between the two circles.}
	\label{Fig1}
\end{figure}

Introduced by Heller and Tomsovic in 1993 \cite{10.1063/1.881358}, the family of lemon billiards has since been the subject of extensive research \cite{bunimovich2016another, 10.1063/1.4850815, PhysRevE.63.056203, PhysRevE.64.016214, PhysRevE.59.303}, including more recent studies \cite{PhysRevE.106.054203, Jin_2021, Bunimovich02102021}. These systems are defined by a boundary formed from the intersection of two identical circles of radius $R$, with their centers separated by a distance $2B$, where $2B<2R$.
The circles are positioned symmetrically along the horzinotal-axis:
 one is shifted left by $B$, the other right by $B$, so the centers of the circles lie at $x=\pm B$, with $B \in [0,1)$. Figure \ref{Fig1} illustrates this configuration: the blue and red curves represent the left- and right-shifted circles, respectively, and the cyan region highlights their intersection, which defines the billiard domain.

The boundary of the billiard is described implicitly in Cartesian coordinates by the equations:

\[
\begin{array}{ll}
	(x + B)^2 + y^2 = R^2, \quad x > 0,
	\\
	(x - B)^2 + y^2 = R^2, \quad x < 0.
\end{array}
\]

The two circles intersect at the points $(0, \pm \sqrt{R^2-B^2})$, marked by green dots in Figure \ref{Fig1}. The total length of the billiard boundary is given by:
$$ \mathcal{L} = 2 R \arcsin\left( \dfrac{\sqrt{R^2 - B^2}}{R}\right) ,$$
and a complete derivation is provided in Appendix A.

The dynamics of the system is characterized by the pair $(s,\alpha)$), where $s$ is the arc-length along the boundary, measured counterclockwise from the point $(0, -\sqrt{R^2-B^2})$, and $\alpha \in (0, \pi)$ is the angle formed between the trajectory of the particle and the tangent to the boundary at the point of collision, measured counterclockwise.

\begin{figure}[!htb]
	\includegraphics[width=\linewidth]{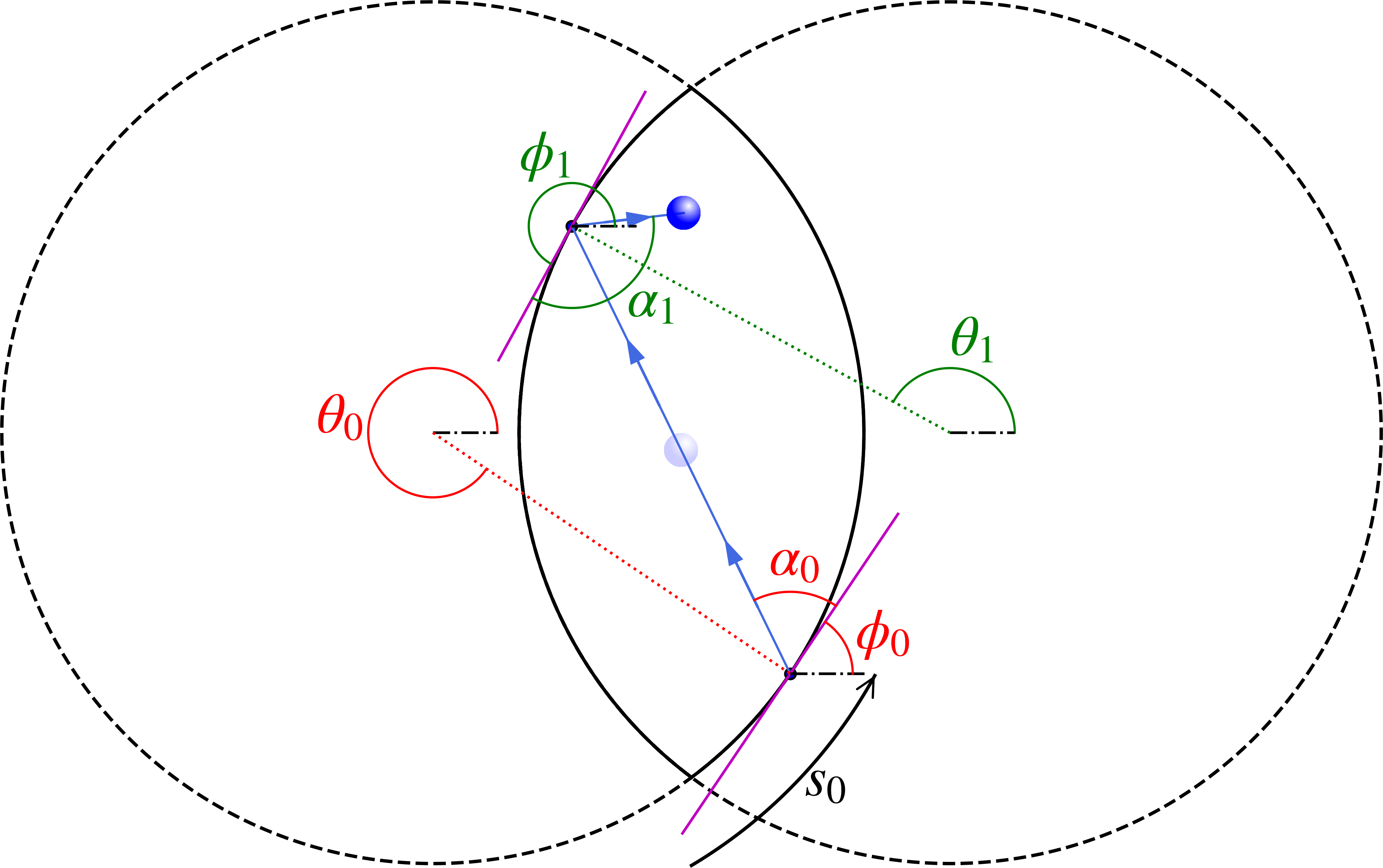}
	\caption{(Color online): An illustration of two consecutive collisions of a particle and the angles involved in the billiard.
	}
	\label{Fig2}
\end{figure}

Figure \ref{Fig2} presents a schematic of the billiard motion and illustrates the definition of these angles at two successive collisions. The red angle represents the initial collision parameters, while the green angle shows those after the subsequent collision. Since no external forces or potentials act inside the billiard, the particle moves along straight-line segments at constant speed between collisions. These trajectories are shown as blue lines.

To describe the system's evolution, we begin with a particle initially located at arc-length $s_n$ with initial angle of injection $\alpha$. The particle's Cartesian coordinates $(x_n, y_n)$ can be obtained using the geometry of the billiard:

\begin{equation}
\begin{array}{ll}
	y_n =\begin{cases}
		R \sin\left( \dfrac{s_n}{R} + \arcsin\left( \dfrac{-\sqrt{1 - B^2}}{R} \right) \right), 
		& \text{if } s_n \leq \dfrac{L}{2}, \\[3ex]
		- R \sin\left( \dfrac{s_n - \frac{L}{2}}{R} - \arcsin\left( \dfrac{\sqrt{1 - B^2}}{R} \right) \right),	& \text{if } s_n > \dfrac{L}{2}.
	\end{cases}  \vspace{1cm} \\
	x_n =\begin{cases}
		-B + \sqrt{R^2 - y_n^2}
		& \text{if } s_n \leq \dfrac{L}{2}, \\[2ex]
		B - \sqrt{R^2 - y_n^2}
		& \text{if } s_n > \dfrac{L}{2}.
	\end{cases}
\end{array}
\end{equation}

To determine the global direction of the particle, in addition to the angle $\alpha_n$, it is necessary to define the slope $\phi_n$ of the tangent line, measured counterclockwise from the horizontal axis. To do this, we introduce the polar angle $\theta_n$, defined as the angle (in standard position, also measured counterclockwise from the horizontal axis) between the center of the arc on which the particle lies and its position $(x_n,y_n$). Figure \ref{Fig2} illustrates these angles to provide a clearer visualization. The $\theta_n$ angle is given by:

$$\theta_n = \left\{\begin{array}{ll}
	\arctan(\dfrac{y_n}{ x_n + B}), \quad \text{if } x_n >0, \\[2ex]
	\arctan(\dfrac{y_n}{ x_n - B}), \quad \text{if } x_n \leq 0
\end{array} \right.
$$
From this, the Cartesian coordinates of the position $(x_n,y_n)$ can be expressed in terms of $\theta_n$ as:
\begin{equation}
	\begin{aligned}
		x(\theta_n) \equiv x_n &= R\cos\theta_n,\\
		y(\theta_n) \equiv y_n &= R\sin\theta_n.\\
	\end{aligned}
\end{equation}
and then the angle $\phi_n$ is given by
\begin{equation}
	\phi_n = \arctan\qty[\frac{y'\qty(\theta_n)}{x'\qty(\theta_n)}]\mod{2\pi},
\end{equation}
where the prime indicates the derivative with respect to $\theta_n$.
 
Finally, the global direction of motion 
$\mu$, measured counterclockwise from the horizontal axis, is then obtained by combining the local angle of incidence $\alpha_n$ with the tangent angle $\phi_n$:
\begin{equation}
	\mu_n = (\alpha_n + \phi_n) \mod{2\pi}.
\end{equation}
 
\begin{figure}[!htb]
 	\includegraphics[width=\linewidth]{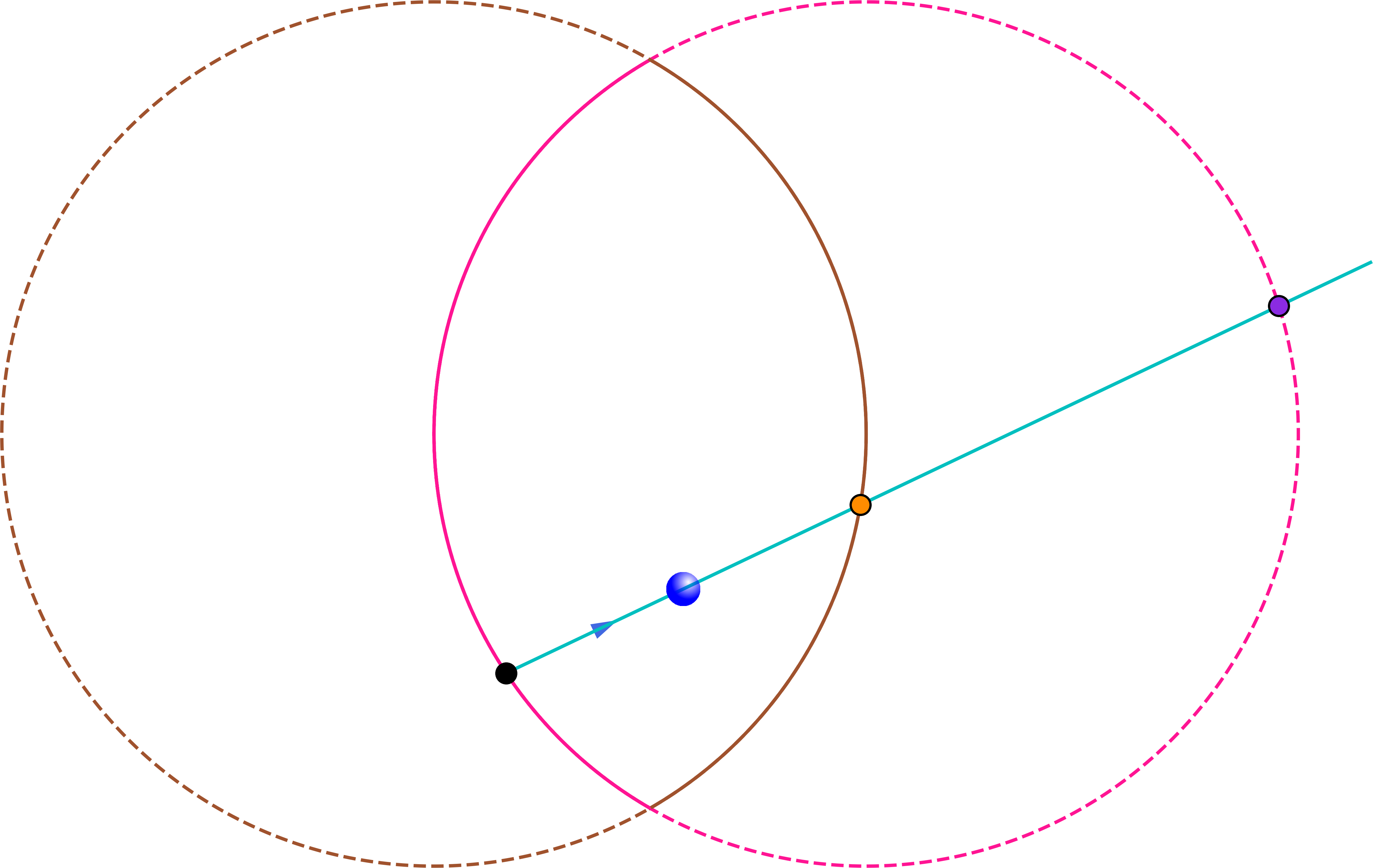}
 	\caption{Schematic representation of the calculation of the time to the next collision. From the particle’s current position (black dot), an extended line is constructed along its trajectory (cyan line), intersecting both the left- and right-shifted circles. The intersection points are marked by orange and purple dots. The smaller of the two corresponding time intervals determines the effective time at which the next collision occurs.}
 	\label{Fig3}
 \end{figure}

 Since no forces act on the particle during free motion, its trajectory between collisions follows a straight line, parameterized by
\begin{equation}
	\label{eq:collisions}
	\begin{aligned}
		x_{n + 1} &= x_n + v_n\cos\qty(\mu_n)\Delta{t},\\
		y_{n + 1} &= y_n + v_n\sin\qty(\mu_n)\Delta{t},
	\end{aligned}
\end{equation}
where $\Delta{t}$ is the time interval between successive collisions. To determine $\Delta t$, we compute the times $\Delta t_+$ and $\Delta t_-$, which represent the time it takes for a particle initially at $\qty(x_n, y_n)$, moving in the global direction $\mu_n$, to reach the circles whose centers are shifted to the left and right by a distance $B$, respectively. Figure~\ref{Fig3} illustrates the determination of $\Delta t_+$ and $\Delta t_-$. From the particle’s current position (marked by a black dot), its trajectory is extended until it intersects the right-shifted circle (shown in pink) and the left-shifted circle (shown in brown). This line is illustrated in cyan. The intersection with the left-shifted circle is marked by a purple dot, and the corresponding time interval is denoted by $\Delta t_+$. Similarly, the intersection with the right-shifted circle is marked by an orange dot, with the respective time interval denoted by $\Delta t_-$. These intersections points are obtained by solving  
$$(x_{n+1}+B)^2 + y_{n+1}^2 = R^2$$
for $\Delta t_+$, and
$$(x_{n+1}-B)^2 + y_{n+1}^2 = R^2$$
for $\Delta t_-$, where $x_{n+1}$ and $y_{n+1}$ are given by Eqs.~\eqref{eq:collisions}.
As a result, the time it takes for the particle to reach these points is determined by solving the corresponding quadratic equations. For $\Delta t_+$:
\begin{equation}
	\begin{aligned}
		\qty(\Delta{t}_+)^2 + 2 v_n\qty[(x_n+B)\cos\mu_n + y_n\sin\mu_n]\Delta{t}_+ + \\
		+ (x_n + B)^2 + y_n^2 - R^2 = 0,
	\end{aligned}
\end{equation}
and for $\Delta{t}_-$:
\begin{equation}
	\begin{aligned}
		\qty(\Delta{t}_-)^2 + 2 v_n\qty[(x_n-B)\cos\mu_n + y_n\sin\mu_n]\Delta{t}_- + \\
		+ (x_n - B)^2 + y_n^2 - R^2 = 0,
	\end{aligned}
\end{equation}
The solutions (considering only non-negative times) are given by:
\begin{equation}
	\Delta{t}_\pm = \frac{-b_\pm + \sqrt{b_\pm^2 - 4c_\pm}}{2},
\end{equation}
where
\begin{equation}
	\begin{aligned}
		b_+ &= 2\qty[(x_n+B)\cos\mu_n + y_n\sin\mu_n], \\ 
		b_- &= 2\qty[(x_n-B)\cos\mu_n + y_n\sin\mu_n], \\
		c_+ &= (x_n + B)^2 + y_n^2 - R^2, \\
		c_- &= (x_n - B)^2 + y_n^2 - R^2.
	\end{aligned}
\end{equation}
Then, the time interval until the next collision, $\Delta{t}$, is given by the smaller of the two values:
 $$\Delta{t} = \min(\Delta{t}_+, \Delta{t}_-).$$
 
Finally, the new arc-length position $s_{n+1}$ where the particle hits the boundary, is given by
 {\small
 \begin{equation}
 	s_{n+1} =
 	\begin{cases}
 		R \left( \arcsin\left( \frac{y_{n+1}}{R} \right) - \arcsin\left( \frac{-\sqrt{R^2 - B^2}}{R} \right) \right),
 		& \text{if } x_{n+1} \geq 0, \\[2ex]
 		\dfrac{L}{2} + R \left( -\arcsin\left( \frac{y_{n+1}}{R} \right) + \arcsin\left( \frac{\sqrt{R^2 - B^2}}{R} \right) \right),
 		& \text{if } x_{n+1} < 0,
 	\end{cases}
 \end{equation}
}%
\noindent and the direction of the particle immediately after the collision is determined by
\begin{equation}
	\alpha_{n + 1} = (\phi_{n + 1} - \mu_n) \mod{\pi}.
\end{equation}

\begin{figure*}
	\includegraphics[width=\linewidth]{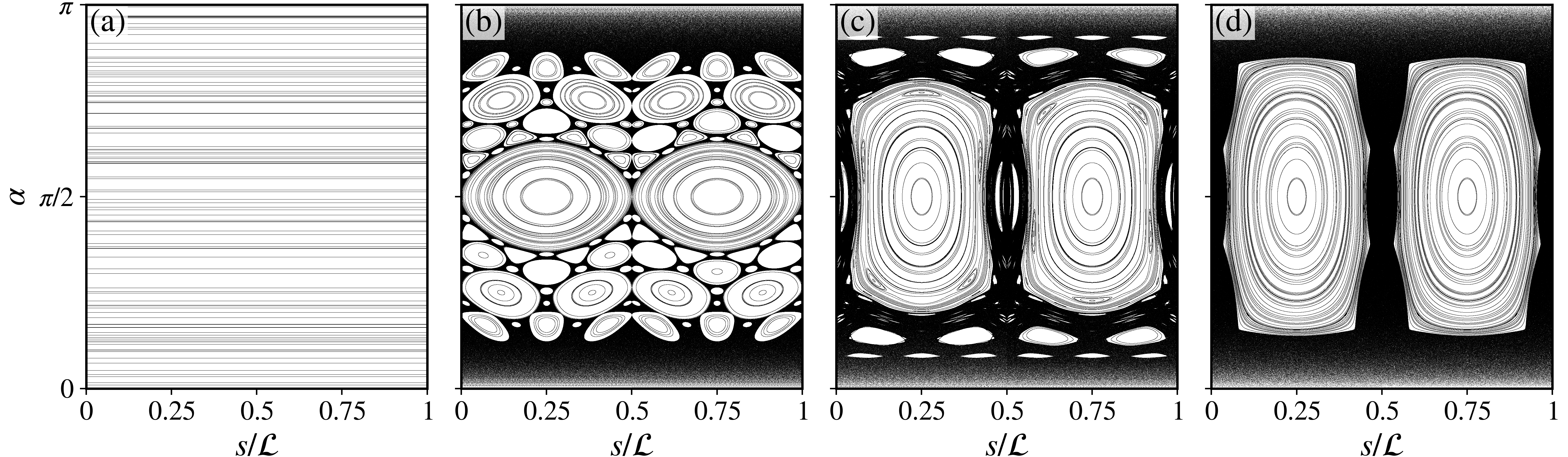}
	\caption{Phase space structure for different values of the parameter $B$.
		(a) For $ B=0 $, the billiard is a perfect circle and the system is integrable; phase space is filled with invariant curves.
		(b) For $ B=0.1 $, integrability breaks, and chaotic regions emerge alongside stability islands.
		(c) For $ B=0.7 $, the central period-2 island becomes dominant.
		(d) For $ B=0.9 $, the phase space is mostly chaotic, with a single large regular island.}
	\label{Fig4}
\end{figure*}

The structure of the phase space is shown in Fig.~\ref{Fig4}. We normalize the arclength $s$ by dividing it by the total boundary length $\mathcal{L}$ to make the phase space dimensionless and comparable. Throughout this work, we consider $R=1$ without loss of generality. For $B=0$, the boundary reduces to a perfect circle, and the system is integrable. As a result, the phase space is filled with straight lines corresponding to regular trajectories, as shown in Fig.~\ref{Fig4}(a). As $B$ increases, integrability breaks down, and several islands emerge, as illustrated for $B=0.1$ in Fig.~\ref{Fig4}(b). With further increase of $B$, the large regular island around the period-2 orbit grows, while the smaller islands shrink and eventually disappear, leaving only the dominant period-2 island embedded in a fully chaotic sea. This evolution is shown for $B=0.7$ and $B=0.9$ in Fig.~\ref{Fig4}(c) and Fig.~\ref{Fig4}(d), respectively.

\section{Survival Probability} \label{sec:survivalprobability}

A key quantity used to investigate the statistical behavior of transport phenomena is the survival probability, which reflects the chance that a particle remains in a certain region despite having an escape option. In this section, we explore the escape dynamics of particles through a hole of size $h$, measured in arc-length units, positioned along the boundary of the billiard.

The survival probability quantifies the likelihood that a particle remains inside the system after $n$ interactions. In other words, it corresponds to the fraction of particles that have not escaped by the $n$th iteration. It can be calculated using the following expression:
\begin{equation}
	\label{eq:Survivel_prob}
	P(n) = \frac{N_{\text{surv}}(n)}{M},
\end{equation}
where $M$ is the total number of initial conditions (particles), and $N_{\text{surv}}$ is the number of particles that remain in the billiard up to the $n$th iteration. 

To introduce escape dynamics into the billiard system, we consider two exits (activating only one at a time) placed along the boundary. Each hole has size $h$, measured in arc-length units. These exits are located at $s_{exit} = 0.50 \mathcal{L}$, situated in a fully chaotic region, and $s_{exit} = 0.75 \mathcal{L}$, lying in a mixed region with both chaotic and regular dynamics.

The methodology involves initializing an ensemble of $M =10^6$ particles randomly distributed across the entire phase space and tracking them for up to $N=10^6$ boundary collisions, or until they escape through the active hole, which occurs when a particle hits the open exit. The statistical analysis is performed based on the number of collisions each particle undergoes before escaping. For each iteration $n$, the number of remaining particles $N_{\text{surv}}(n)$ inside the billiard is recorded. This data allows us to compute the survival probability for each iteration $n$, thereby enabling a statistical analysis of the escape dynamics.

\begin{figure}[!htb]
	\includegraphics[width=\linewidth]{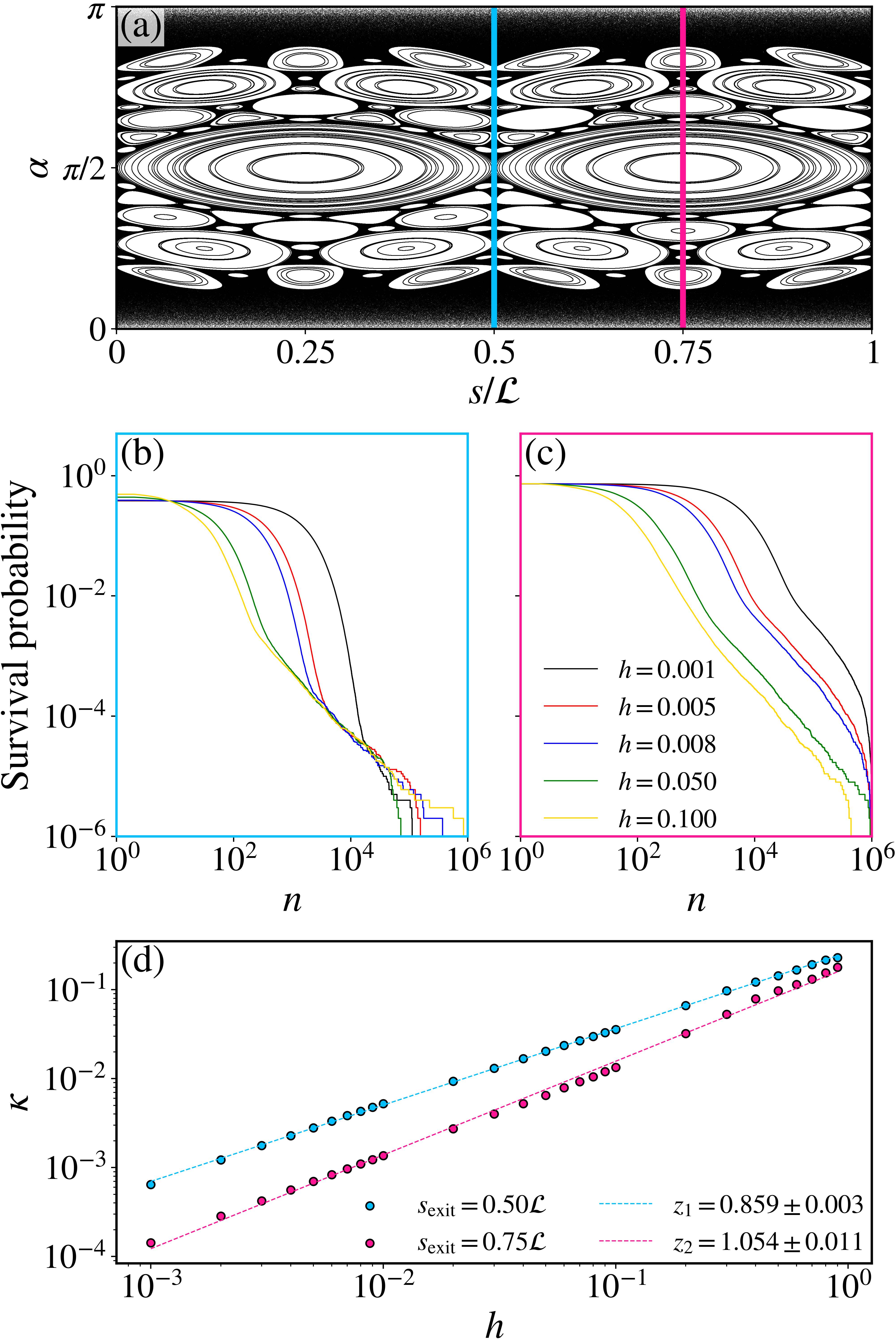}
	\caption{(Color online): 
		(a) Phase space for \( B = 0.1 \), with hole positions indicated at \( s_{\text{exit}} = 0.50\mathcal{L} \) (chaotic region) and \( s_{\text{exit}} = 0.75\mathcal{L} \) (mixed region).
		(b) Survival probability curves for different hole sizes \( h \) at \( s_{\text{exit}} = 0.50\mathcal{L} \).
		(c) Survival probability curves for different \( h \) at \( s_{\text{exit}} = 0.75\mathcal{L} \).
		(d) Log-log plot of the escape rate \( \kappa \) as a function of \( h \), showing power-law scaling \( \kappa(h) \sim h^z \) for both hole positions.
	}
	\label{Fig5}
\end{figure}

We start our analysis considering $B=0.1$ and varies the size of the exit $h$. The Fig.~\ref{Fig5}(a) shows the space phase for this configuration showing also the position of the two exits. The survival probability for the exits located in $s_{exit} = 0.50 \mathcal{L}$ and $s_{exit} = 0.75 \mathcal{L}$ are shown in Fig.~\ref{Fig5}~(b) and (c), respectively, for five different exit sizes $h$. It is possible to see that we have similiar behavior for either holes, which is the compostion of 2 patterns, for sort times,  the survival probability decays exponentially as
\begin{equation}
	\label{eq:pn_exp}
	P(n) \sim \exp\qty(-\kappa n)
\end{equation}
where $\kappa > 0$ is the escape rate, which is commom for system the present chaotic behavior ~\cite{BORIN2023113965,LEONEL20121669}, while for longer times, a power-law tail emerges, described by
\begin{equation}
	\label{eq:pn_pl}
	P(n) \sim A n^{-\gamma}
\end{equation}
where $A$ is also a non-negative constant and $\gamma$ is the power law decay rate, which is a characteristic feature of the stickiness effect~\cite{Altmann2009,Livorati2012}. Due to the stickiness effect, particles might be trapped near stability islands and resonance zones for a long, but finite, time leading to long escape times and causing the aforementioned deviations from the exponential decay.

The behavior of the exponential decay at short times is similar for both holes, differing only in the values of the decay rate. We investigate how the decay rate 
$\kappa$ depends on the hole size $h$, and the results are shown in Figure~\ref{Fig5}(c). We find that this dependence follows a power law over several orders of magnitude: $\kappa(h) \sim h^z$, where $z$ is the scaling exponent. The values obtained are $z_1 = 0.859$ and $z_2 = 1.054$ for holes located at $s_{\text{exit}} = 0.50 L$ and $s_{\text{exit}} = 0.75 L$, respectively. This shows that the decay rate exhibits scaling invariance with respect to variations in $h$  \cite{PhysRevE.69.066218, 10.1063/5.0219961, CZAJKOWSKI2024114994} 

An interesting difference emerges when comparing the power-law tail behavior for holes located in chaotic regions versus those in mixed phase space regions. For a hole placed in a chaotic region, the survival probability exhibits a nearly universal power-law tail, independent of the hole size $h$, converging to the same behavior. In contrast, when the hole is located in regions containing stability islands, the power-law decay is still present but shifted.

This difference arises mainly because placing the hole partially or entirely over an island can eliminate all orbits near that island. In chaotic regions, the stickiness phenomenon remains statistically similar across different hole sizes. However, when the hole covers regions with islands, some sticky regions and resonance zones may be destroyed, affecting the power-law decay. Therefore, our results suggest that escape is faster when the hole is placed in chaotic regions without stability islands, consistent with findings in previous studies \cite{SALES_10.1063/5.0222215, HANSEN2018355, HANSEN20163634}.

\begin{figure}[!htb]
	\includegraphics[width=\linewidth]{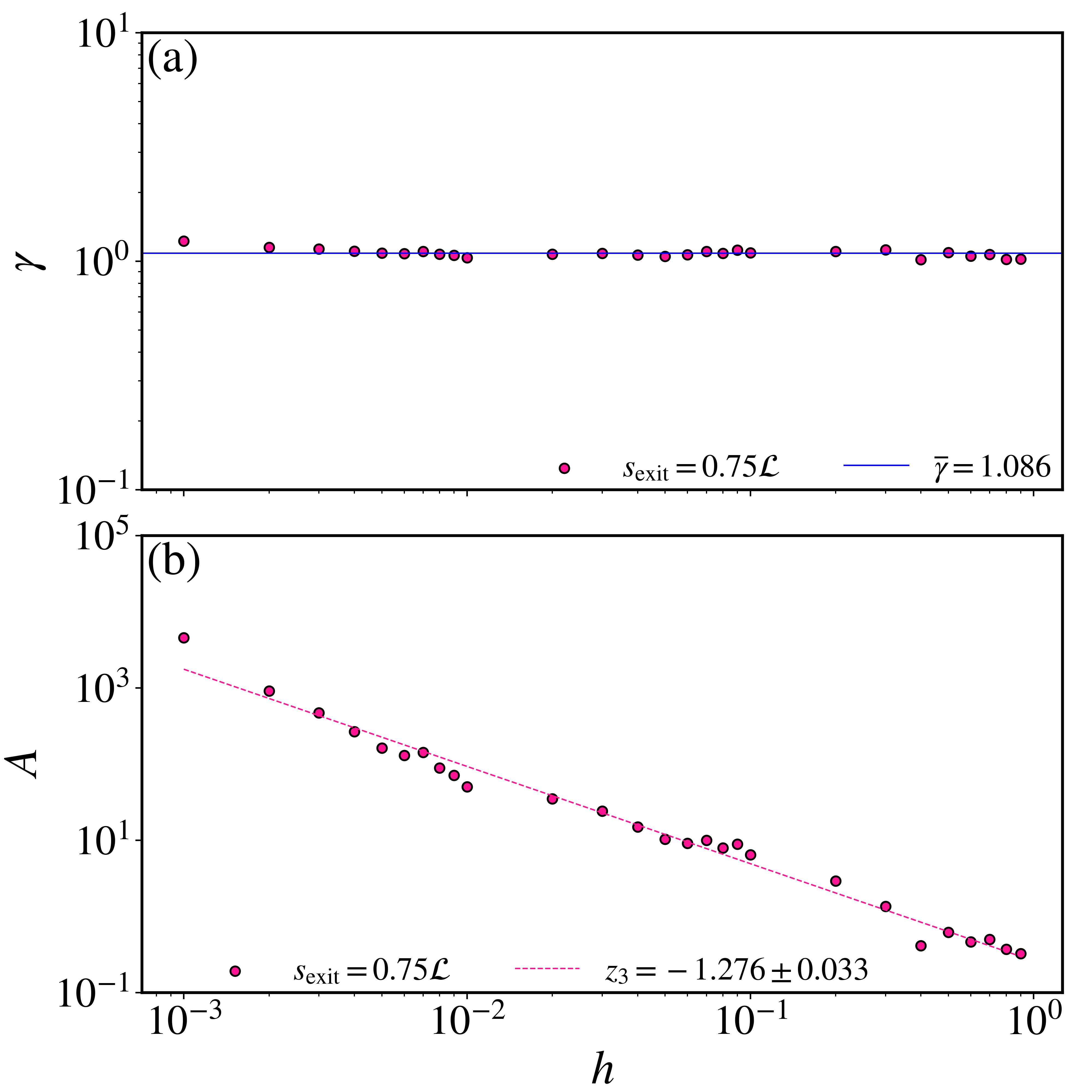}
	    \caption{(Color online): 
		Scaling analysis of the power-law tail when the hole is placed at \( s_{\text{exit}} = 0.75\mathcal{L} \).
		(a) Decay exponent \( \gamma \) as a function of \( h \), indicating a constant behavior with average \( \bar{\gamma} = 1.086 \).
		(b) Coefficient \( A \) of the power-law decay versus \( h \), exhibiting a power-law scaling \( A(h) \sim h^{z_3} \).
	}
	\label{Fig6}
\end{figure}

We further analyze the different patterns in the power-law tail for the survival probability in the case where the hole is located at $s_{\text{exit}} = 0.75 L$, i.e., over regions with islands. By fitting this region with Eq.~\eqref{eq:pn_pl}, we study how the parameters $\gamma$ and $A$ vary with $h$. The analysis of $\gamma$ and $A$ as functions of $h$ [Figs.~\ref{Fig6}(a) and \ref{Fig6}(b), respectively] reveals that the decay exponent $\gamma$ is essentially constant, with a mean value $\bar{\gamma} = 1.086$, while $A$, similarly to $\kappa$, shows power-law scaling: $A(h) \sim h^{z_3},$ with $z_3 = -1.276 \pm 0.033.$

Scaling invariance is a property where a system’s behavior remains unchanged under rescaling of its parameters, revealing underlying structures or universal patterns. Examples include fractional dynamics \cite{BORIN2024114597}, social networks \cite{OLIVEIRA20183376}, and biological systems \cite{PhysRevE.110.054201}. In the case of the lemon billiard, the system depends on two main parameters: the shape parameter $B$ and the size of the hole $h$. We have already observed scaling behavior in the decay rate $\kappa$ and in the coefficient $A$ of the power law tail when varying $h$. Given this result, a natural question arises: Does the system also exhibit scaling invariance when we vary the shape of the billiard, i.e., the parameter $B$? 

\begin{figure}[!htb]
	\includegraphics[width=\linewidth]{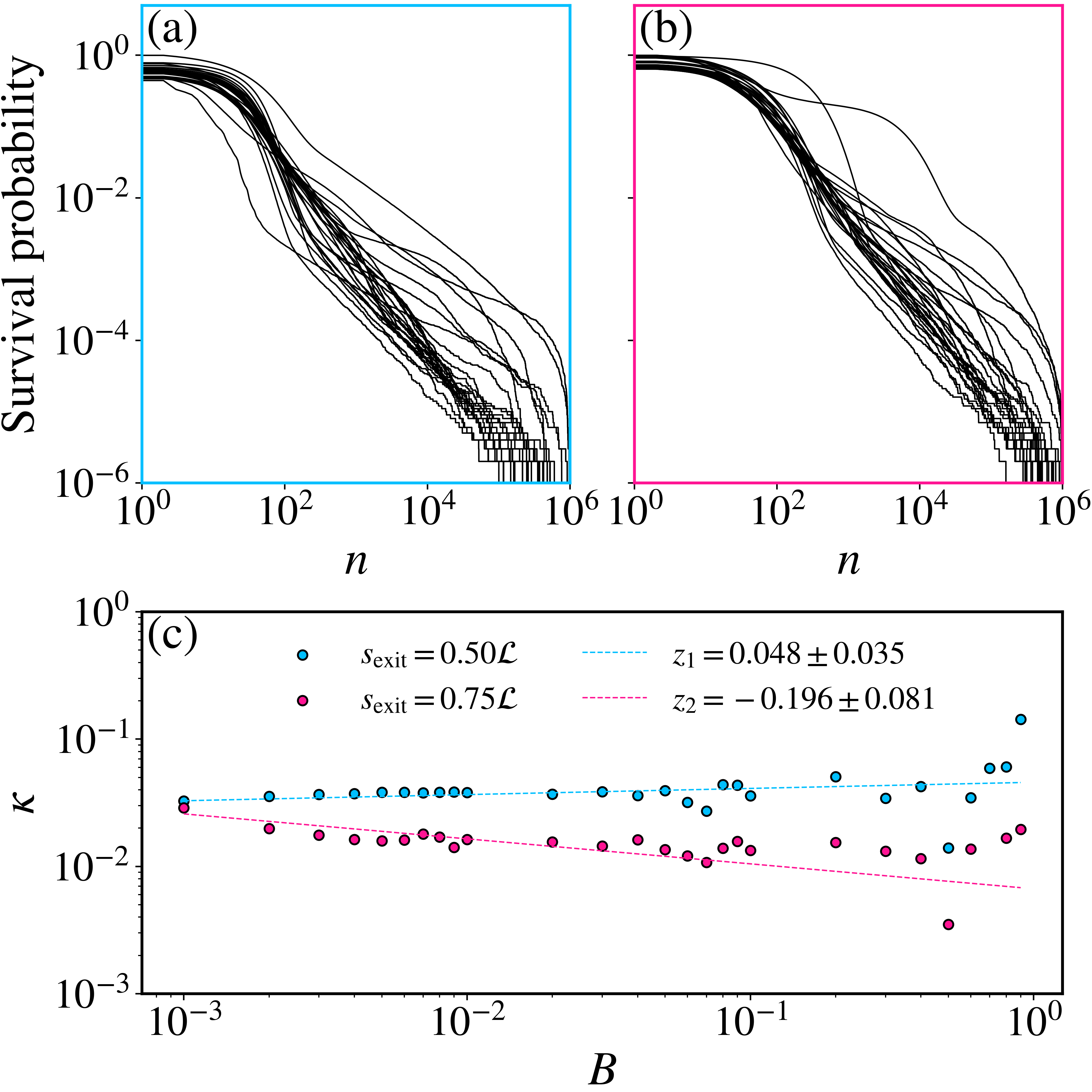}
    \caption{(Color online): 
	Survival probability for fixed hole size \( h = 0.1 \) and different values of the shape parameter \( B \).
	(a) Survival probability for holes at \( s_{\text{exit}} = 0.50\mathcal{L} \).
	(b) Survival probability for holes at \( s_{\text{exit}} = 0.75\mathcal{L} \).
	(c) Escape rate \( \kappa \) as a function of \( B \) for both hole positions. A power-law dependence is observed for small \( B \), which deteriorates as \( B \) increases.
}
	\label{Fig7}
\end{figure}

To address this question, Fig.~\ref{Fig7}(a)-(b) shows the of the survival probability for fixed hole size $h=0.1$ and several values of $B$, with the hole centered at $s_{exit} = 0.50 \mathcal{L}$, and $s_{exit} = 0.75 \mathcal{L}$, respectively. It is possible to observe that the curves exhibit the same general pattern as before: an initial exponential decay followed by a power-law tail. Additionally, we also observe some curves that present a stretched exponential decay~\cite{DETTMANN2012403,deFaria2016,LIVORATI2018225}. Figure \ref{Fig7}(c) shows how the short-time exponential decay rate $\kappa$ varies with $B$ for holes at $s_{exit} = 0.75 \mathcal{L}$ (blue) and $s_{exit} = 0.75 \mathcal{L}$ (pink). For small values of $B$, a power-law behavior is evident, but this behavior deteriorates as $B$ increases.


\section{Conclusion} \label{sec:conclusion}

In this work, we investigated the escape dynamics of the lemon billiard by analyzing the survival probability of particles escaping through specific holes positioned along the boundary, focusing on whether this observable exhibits scaling invariance under variations of the system parameters, namely the hole size $h$ and the shape parameter $B$.

We started by providing a detailed geometric and dynamical description of the lemon billiard, which is formed by the intersection of two identical circles of radius $R$, separated by a distance $2B$. The dynamics were described using arc length and incidence angles, and we developed a framework to simulate the motion of particles as they collide with the billiard boundary.

To analyze the escape dynamics, we introduced a hole of size $h$ at two distinct positions along the boundary: one located in a fully chaotic region ($s_{\text{exit}} = 0.50\mathcal{L}$) and another in a mixed region that includes stability islands ($s_{\text{exit}} = 0.75\mathcal{L}$). For each configuration, we computed the survival probability $P(n)$, defined as the fraction of particles that remain inside the billiard after $n$ collisions. Our results show that the survival probability exhibits two distinct decay regimes: an initial exponential decay at short times, followed by a power-law decay at longer times. The exponential regime is associated with the chaotic dynamics, while the power-law tail results from the stickiness effect caused by particles temporarily trapped near regular islands.

We studied the dependence of the exponential decay rate $\kappa$ on the hole size $h$, and found a power-law scaling $\kappa(h) \sim h^z$, where the exponent $z$ depends on the hole’s position. Additionally, in the mixed region, the amplitude $A$ of the power-law tail also exhibits scaling behavior with respect to $h$, while the power-law exponent $\gamma$ remains approximately constant.

Finally, we extended our analysis by fixing the hole size $h$ and varying the shape parameter $B$. The survival probability continued to show the same general pattern of exponential and power-law decay, although some curves exhibited a stretched exponential decay. We observed that for small values of $B$, the exponential decay rate $\kappa$ again follows a power-law dependence, indicating scaling behavior. However, as $B$ increases, this behavior becomes less clear.

In summary, our study reveals that the survival probability in the lemon billiard exhibits scaling invariance with respect to the hole size $h$, both in the exponential decay regime and in the power-law tail. Scaling invariance with respect to the shape parameter $B$ is also observed, but only within a limited range of values.

\section*{Acknowledgments}
This study was financed, in part, by the São Paulo Research Foundation (FAPESP), Brasil. Process Numbers \#2024/06749-8, \#2022/03612-6, \#2019/14038-6, and \# 2021/09519-5.
E.D.L. acknowledges support from Brazilian agencies CNPq (No. 301318/2019-0, 304398/2023-3) 

\appendix

\section{Calculation of the Billiard Boundary Length} \label{appendix:length}

To compute the total length of the billiard boundary, we start by calculating the arc length of a single circular segment. Due to the geometry of the lemon billiard, the full boundary consists of two symmetric arc segments of equal length. Therefore, it is sufficient to calculate just one arc and then double the result to obtain the total length.

The arc lies on a circle centered at \( (-B, 0) \) with radius \( R \), given implicitly by:
\begin{equation}
	(x + B)^2 + y^2 = R^2. 
	\label{eq:circle}
\end{equation}

We apply the arc length formula for implicitly defined curves:
\begin{equation}
	L = \int_{y_1}^{y_2} \sqrt{1 + \left( \frac{dx}{dy} \right)^2} \, dy.
	\label{eq:arc_length}
\end{equation}

Differentiating equation~\eqref{eq:circle} with respect to \( y \):
\begin{align}
	\frac{d}{dy} \left( (x + B)^2 + y^2 \right) &= 0 \nonumber \\
	2(x + B)\frac{dx}{dy} + 2y &= 0 \nonumber \\
	\Rightarrow \quad \frac{dx}{dy} &= -\frac{y}{x + B}.
	\label{eq:dxdy}
\end{align}

Substituting equation~\eqref{eq:dxdy} into \eqref{eq:arc_length}:
\begin{align}
	L &= \int_{y_1}^{y_2} \sqrt{1 + \left( \frac{y}{x + B} \right)^2} \, dy.
	\label{eq:arc_step1}
\end{align}

From equation~\eqref{eq:circle}, we solve for \( x + B \) in terms of \( y \):
\begin{equation}
	(x + B)^2 = R^2 - y^2.
	\label{eq:xplusb}
\end{equation}

Substituting equation~\eqref{eq:xplusb} into \eqref{eq:arc_step1}:
\begin{align}
	L &= \int_{y_1}^{y_2} \sqrt{1 + \frac{y^2}{R^2 - y^2}} \, dy 
	= \int_{y_1}^{y_2} \sqrt{\frac{R^2}{R^2 - y^2}} \, dy \nonumber \\
	&= \int_{y_1}^{y_2} \frac{R}{\sqrt{R^2 - y^2}} \, dy.
	\label{eq:arc_integral}
\end{align}

The integral in equation~\eqref{eq:arc_integral} evaluates to:
\begin{equation}
	L = R \left[ \arcsin\left( \frac{y}{R} \right) \right]_{y_1}^{y_2}.
	\label{eq:arcsin_eval}
\end{equation}

The circular arcs intersect at \( y = \pm \sqrt{R^2 - B^2} \), so we set \( y_1 = -\sqrt{R^2 - B^2} \) and \( y_2 = \sqrt{R^2 - B^2} \). Then equation~\eqref{eq:arcsin_eval} becomes:
\begin{align}
	L &= R \left[ \arcsin\left( \frac{\sqrt{R^2 - B^2}}{R} \right) - \arcsin\left( \frac{-\sqrt{R^2 - B^2}}{R} \right) \right] \nonumber \\
	&= 2R \arcsin\left( \frac{\sqrt{R^2 - B^2}}{R} \right) 
	\label{eq:single_arc}
\end{align}

Since the billiard boundary consists of two identical arcs, the total boundary length is:
\begin{equation}
	\mathcal{L} = 2L = 4R \arcsin\left( \frac{\sqrt{R^2 - B^2}}{R} \right).
	\label{eq:total_length}
\end{equation}



\subsection*{Comparison with the Literature}

In the work \cite{PhysRevE.103.012204}, the authors give the total boundary length as:
\begin{equation}
	\mathcal{L} = 4 \arctan \left( \sqrt{B^{-2} - 1} \right),
	\label{eq:lozej}
\end{equation}
assuming \( R = 1 \).

To verify that equations~\eqref{eq:total_length} and \eqref{eq:lozej} are equivalent, define:
\begin{equation}
	\theta = \arcsin\left( \sqrt{1 - B^2} \right),
	\label{eq:theta_arcsin}
\end{equation}
so that:
\begin{align}
	\sin \theta &= \sqrt{1 - B^2}, \label{eq:sin_theta} \\
	\cos \theta &= B. \label{eq:cos_theta}
\end{align}

Then the tangent of \( \theta \) is:
\begin{equation}
	\tan \theta = \frac{\sin \theta}{\cos \theta} = \frac{\sqrt{1 - B^2}}{B}.
	\label{eq:tan_theta}
\end{equation}

Now define:
\begin{equation}
	\phi = \arctan\left( \sqrt{B^{-2} - 1} \right) = \arctan\left( \frac{\sqrt{1 - B^2}}{B} \right).
	\label{eq:phi_arctan}
\end{equation}

From equations~\eqref{eq:tan_theta} and \eqref{eq:phi_arctan}, it follows that \( \theta = \phi \), and hence:
\begin{equation}
	\arcsin\left( \sqrt{1 - B^2} \right) = \arctan\left( \sqrt{B^{-2} - 1} \right).
	\label{eq:identity}
\end{equation}

Therefore, both expressions for \( \mathcal{L} \), equations~\eqref{eq:total_length} and \eqref{eq:lozej}, are equivalent when \( R = 1 \), and the identity confirms the consistency between our geometric derivation and the formula found in the literature.

\bibliography{References_SurvProb_LemonBilliard}

@PREAMBLE{
 "\providecommand{\noopsort}[1]{}" 
 # "\providecommand{\singleletter}[1]{#1}%" 
}

@book{chernov2006chaotic,
  title={Chaotic Billiards},
  author={Chernov, N. and Markarian, R.},
  isbn={9780821840962},
  lccn={2006042819},
  series={Mathematical surveys and monographs},
  year={2006},
  publisher={American Mathematical Society}
}

@article{boucher2025buzz,
  title={Buzz pollination: investigations of pollen expulsion using the discrete element method},
  author={Boucher-Bergstedt, Caelen and Jankauski, Mark and Johnson, Erick},
  journal={Journal of the Royal Society Interface},
  volume={22},
  number={222},
  pages={20240526},
  year={2025},
  publisher={The Royal Society},
  doi = {10.1098/rsif.2024.0526},
  url = {https://doi.org/10.1098/rsif.2024.0526}
}

@article{PhysRevE.110.054201,
  title = {Buzz pollination: A theoretical analysis via scaling invariance},
  author = {Borin, Daniel and de Brito, Vinicius Lourenço Garcia and Leonel, Edson Denis and Hansen, Matheus},
  journal = {Phys. Rev. E},
  volume = {110},
  issue = {5},
  pages = {054201},
  numpages = {6},
  year = {2024},
  month = {Nov},
  publisher = {American Physical Society},
  doi = {10.1103/PhysRevE.110.054201},
  url = {https://link.aps.org/doi/10.1103/PhysRevE.110.054201}
}

@article{de2024analytical,
  title={Analytical methods in celestial mechanics: satellites’ stability and galactic billiards},
  author={De Blasi, Irene},
  journal={Astrophysics and Space Science},
  volume={369},
  number={5},
  pages={52},
  year={2024},
  publisher={Springer},
  doi = {10.1007/s10509-024-04312-8},
  url = {https://doi.org/10.1007/s10509-024-04312-8}
}

@article{Barutello_2023,
doi = {10.1088/1361-6544/acdec2},
url = {https://dx.doi.org/10.1088/1361-6544/acdec2},
year = {2023},
month = {jun},
publisher = {IOP Publishing},
volume = {36},
number = {8},
pages = {4209},
author = {Barutello, Vivina L and De Blasi, Irene and Terracini, Susanna},
title = {Chaotic dynamics in refraction galactic billiards},
journal = {Nonlinearity}
}

@article{PhysRevLett.94.100201,
  title = {Open Circular Billiards and the Riemann Hypothesis},
  author = {Bunimovich, L. A. and Dettmann, C. P.},
  journal = {Phys. Rev. Lett.},
  volume = {94},
  issue = {10},
  pages = {100201},
  numpages = {4},
  year = {2005},
  month = {Mar},
  publisher = {American Physical Society},
  doi = {10.1103/PhysRevLett.94.100201},
  url = {https://link.aps.org/doi/10.1103/PhysRevLett.94.100201}
}

@article{PhysRevLett.132.157101,
  title = {Billiards with Spatial Memory},
  author = {Albers, Thijs and Delnoij, Stijn and Schramma, Nico and Jalaal, Maziyar},
  journal = {Phys. Rev. Lett.},
  volume = {132},
  issue = {15},
  pages = {157101},
  numpages = {7},
  year = {2024},
  month = {Apr},
  publisher = {American Physical Society},
  doi = {10.1103/PhysRevLett.132.157101},
  url = {https://link.aps.org/doi/10.1103/PhysRevLett.132.157101}
}

@article{PhysRevLett.108.064102,
  title = {Destruction of Adiabatic Invariance for Billiards in a Strong Nonuniform Magnetic Field},
  author = {Neishtadt, A. I. and Artemyev, A. V.},
  journal = {Phys. Rev. Lett.},
  volume = {108},
  issue = {6},
  pages = {064102},
  numpages = {4},
  year = {2012},
  month = {Feb},
  publisher = {American Physical Society},
  doi = {10.1103/PhysRevLett.108.064102},
  url = {https://link.aps.org/doi/10.1103/PhysRevLett.108.064102}
}

@article{PhysRevLett.68.2867,
  title = {Experimental determination of billiard wave functions},
  author = {Stein, J. and St\"ockmann, H.-J.},
  journal = {Phys. Rev. Lett.},
  volume = {68},
  issue = {19},
  pages = {2867--2870},
  numpages = {0},
  year = {1992},
  month = {May},
  publisher = {American Physical Society},
  doi = {10.1103/PhysRevLett.68.2867},
  url = {https://link.aps.org/doi/10.1103/PhysRevLett.68.2867}
}

@article{PhysRevLett.85.2478,
  title = {Observation of Resonance Trapping in an Open Microwave Cavity},
  author = {Persson, E. and Rotter, I. and St\"ockmann, H.-J. and Barth, M.},
  journal = {Phys. Rev. Lett.},
  volume = {85},
  issue = {12},
  pages = {2478--2481},
  numpages = {0},
  year = {2000},
  month = {Sep},
  publisher = {American Physical Society},
  doi = {10.1103/PhysRevLett.85.2478},
  url = {https://link.aps.org/doi/10.1103/PhysRevLett.85.2478}
}

@article{PhysRevLett.134.188303,
  title = {Universal Law for the Dispersal of Motile Microorganisms in Porous Media},
  author = {Pietrangeli, T. and Foffi, R. and Stocker, R. and Ybert, C. and Cottin-Bizonne, C. and Detcheverry, F.},
  journal = {Phys. Rev. Lett.},
  volume = {134},
  issue = {18},
  pages = {188303},
  numpages = {7},
  year = {2025},
  month = {May},
  publisher = {American Physical Society},
  doi = {10.1103/PhysRevLett.134.188303},
  url = {https://link.aps.org/doi/10.1103/PhysRevLett.134.188303}
}

@article{Jonathan_P_Bird_1999,
doi = {10.1088/0953-8984/11/38/201},
url = {https://dx.doi.org/10.1088/0953-8984/11/38/201},
year = {1999},
month = {sep},
publisher = {},
volume = {11},
number = {38},
pages = {R413},
author = {Jonathan P Bird},
title = {Recent experimental studies of electron transport in open 
quantum dots},
journal = {Journal of Physics: Condensed Matter}
}

@article{bunimovich1974ergodic,
  title={On ergodic properties of certain billiards},
  author={Bunimovich, Leonid A},
  journal={Functional Analysis and Its Applications},
  volume={8},
  number={3},
  pages={254--255},
  year={1974},
  publisher={Springer},
  url = {https://doi.org/10.1007/BF01075700},
  doi = {10.1007/BF01075700} ,
}

@article{PhysRevE.103.012204,
  title = {Effects of stickiness in the classical and quantum ergodic lemon billiard},
  author = {Lozej, Crt and Lukman, Dragan and Robnik, Marko},
  journal = {Phys. Rev. E},
  volume = {103},
  issue = {1},
  pages = {012204},
  numpages = {12},
  year = {2021},
  month = {Jan},
  publisher = {American Physical Society},
  doi = {10.1103/PhysRevE.103.012204},
  url = {https://link.aps.org/doi/10.1103/PhysRevE.103.012204}
}

@article{PhysRevE.75.046204,
  title = {Leaking billiards},
  author = {Nagler, Jan and Krieger, Moritz and Linke, Marco and Sch\"onke, Johannes and Wiersig, Jan},
  journal = {Phys. Rev. E},
  volume = {75},
  issue = {4},
  pages = {046204},
  numpages = {7},
  year = {2007},
  month = {Apr},
  publisher = {American Physical Society},
  doi = {10.1103/PhysRevE.75.046204},
  url = {https://link.aps.org/doi/10.1103/PhysRevE.75.046204},
}

@article{10.1063/1.881358,
    author = {Heller, Eric J. and Tomsovic, Steven},
    title = {Postmodern Quantum Mechanics},
    journal = {Physics Today},
    volume = {46},
    number = {7},
    pages = {38-46},
    year = {1993},
    month = {07},
    issn = {0031-9228},
    doi = {10.1063/1.881358},
    url = {https://doi.org/10.1063/1.881358},
}

@article{bunimovich2016another,
  title={On another edge of defocusing: hyperbolicity of asymmetric lemon billiards},
  author={Bunimovich, Leonid and Zhang, Hong-Kun and Zhang, Pengfei},
  journal={Communications in Mathematical Physics},
  volume={341},
  number={3},
  pages={781--803},
  year={2016},
  publisher={Springer},
  doi = {10.1007/s00220-015-2539-x},
  url = {https://doi.org/10.1007/s00220-015-2539-x},
}

@article{10.1063/1.4850815,
    author = {Chen, Jingyu and Mohr, Luke and Zhang, Hong-Kun and Zhang, Pengfei},
    title = {Ergodicity of the generalized lemon billiards},
    journal = {Chaos: An Interdisciplinary Journal of Nonlinear Science},
    volume = {23},
    number = {4},
    pages = {043137},
    year = {2013},
    month = {12},
    issn = {1054-1500},
    doi = {10.1063/1.4850815},
    url = {https://doi.org/10.1063/1.4850815},
}

@article{PhysRevE.63.056203,
  title = {Quantum-classical correspondences of the Berry-Robnik parameter through bifurcations in lemon billiard systems},
  author = {Makino, H. and Harayama, T. and Aizawa, Y.},
  journal = {Phys. Rev. E},
  volume = {63},
  issue = {5},
  pages = {056203},
  numpages = {11},
  year = {2001},
  month = {Apr},
  publisher = {American Physical Society},
  doi = {10.1103/PhysRevE.63.056203},
  url = {https://link.aps.org/doi/10.1103/PhysRevE.63.056203}
}

@article{PhysRevE.64.016214,
  title = {Chaotic behavior in lemon-shaped billiards with elliptical and hyperbolic boundary arcs},
  author = {Lopac, V. and Mrkonji\ifmmode \acute{c}\else \'{c}\fi{}, I. and Radi\ifmmode \acute{c}\else \'{c}\fi{}, D.},
  journal = {Phys. Rev. E},
  volume = {64},
  issue = {1},
  pages = {016214},
  numpages = {8},
  year = {2001},
  month = {Jun},
  publisher = {American Physical Society},
  doi = {10.1103/PhysRevE.64.016214},
  url = {https://link.aps.org/doi/10.1103/PhysRevE.64.016214}
}

@article{PhysRevE.59.303,
  title = {Classical and quantum chaos in the generalized parabolic lemon-shaped billiard},
  author = {Lopac, V. and Mrkonji\ifmmode \acute{c}\else \'{c}\fi{}, I. and Radi\ifmmode \acute{c}\else \'{c}\fi{}, D.},
  journal = {Phys. Rev. E},
  volume = {59},
  issue = {1},
  pages = {303--311},
  numpages = {0},
  year = {1999},
  month = {Jan},
  publisher = {American Physical Society},
  doi = {10.1103/PhysRevE.59.303},
  url = {https://link.aps.org/doi/10.1103/PhysRevE.59.303}
}

@article{PhysRevE.106.054203,
  title = {Phenomenology of quantum eigenstates in mixed-type systems: Lemon billiards with complex phase space structure},
  author = {Lozej, Crt and Lukman, Dragan and Robnik, Marko},
  journal = {Phys. Rev. E},
  volume = {106},
  issue = {5},
  pages = {054203},
  numpages = {15},
  year = {2022},
  month = {Nov},
  publisher = {American Physical Society},
  doi = {10.1103/PhysRevE.106.054203},
  url = {https://link.aps.org/doi/10.1103/PhysRevE.106.054203}
}

@article{Jin_2021,
doi = {10.1088/1361-6544/abaff2},
url = {https://dx.doi.org/10.1088/1361-6544/abaff2},
year = {2020},
month = {nov},
publisher = {IOP Publishing},
volume = {34},
number = {1},
pages = {92},
author = {Jin, Xin and Zhang, Pengfei},
title = {Hyperbolicity of asymmetric lemon billiards},
journal = {Nonlinearity}
}

@article{Bunimovich02102021,
author = {Leonid A. Bunimovich and Giulio Casati and Tomaž Prosen and Gregor Vidmar},
title = {Few Islands Approximation of Hamiltonian System with divided Phase Space},
journal = {Experimental Mathematics},
volume = {30},
number = {4},
pages = {459--468},
year = {2021},
publisher = {Taylor \& Francis},
doi = {10.1080/10586458.2018.1559777},
URL = { https://doi.org/10.1080/10586458.2018.1559777},
}

@article{LEONEL20121669,
title = {Recurrence of particles in static and time varying oval billiards},
journal = {Physics Letters A},
volume = {376},
number = {20},
pages = {1669-1674},
year = {2012},
issn = {0375-9601},
doi = {https://doi.org/10.1016/j.physleta.2012.03.056},
url = {https://www.sciencedirect.com/science/article/pii/S0375960112004045},
author = {Edson D. Leonel and Carl P. Dettmann}
}

@article{BORIN2023113965,
title = {An investigation of the survival probability for chaotic diffusion in a family of discrete {H}amiltonian mappings},
journal = {Chaos, Solitons \& Fractals},
volume = {175},
pages = {113965},
year = {2023},
issn = {0960-0779},
doi = {https://doi.org/10.1016/j.chaos.2023.113965},
url = {https://www.sciencedirect.com/science/article/pii/S0960077923008664},
author = {Daniel Borin and André Luís Prando Livorati and Edson Denis Leonel}
}

@article{Altmann2009,
title = {Poincar\'e recurrences and transient chaos in systems with leaks},
author = {Altmann, Eduardo G. and T\'el, Tam\'as},
journal = {Phys. Rev. E},
volume = {79},
issue = {1},
pages = {016204},
numpages = {12},
year = {2009},
month = {Jan},
publisher = {American Physical Society},
doi = {10.1103/PhysRevE.79.016204},
url = {https://link.aps.org/doi/10.1103/PhysRevE.79.016204}
}

@article{Livorati2012,
title = {Stickiness in a bouncer model: A slowing mechanism for {F}ermi acceleration},
author = {Livorati, Andr\'e L. P. and Kroetz, Tiago and Dettmann, Carl P. and Caldas, Iber\^e Luiz and Leonel, Edson D.},
journal = {Phys. Rev. E},
volume = {86},
issue = {3},
pages = {036203},
numpages = {9},
year = {2012},
month = {Sep},
publisher = {American Physical Society},
doi = {10.1103/PhysRevE.86.036203},
url = {https://link.aps.org/doi/10.1103/PhysRevE.86.036203}
}

@article{PhysRevE.69.066218,
  title = {Crash test for the Copenhagen problem},
  author = {Nagler, Jan},
  journal = {Phys. Rev. E},
  volume = {69},
  issue = {6},
  pages = {066218},
  numpages = {6},
  year = {2004},
  month = {Jun},
  publisher = {American Physical Society},
  doi = {10.1103/PhysRevE.69.066218},
  url = {https://link.aps.org/doi/10.1103/PhysRevE.69.066218}
}

@article{10.1063/5.0219961,
    author = {Czajkowski, Bruno M. and Viana, Ricardo L.},
    title = {Riddled basins of chaotic synchronization and unstable dimension variability in coupled Lorenz-like systems},
    journal = {Chaos: An Interdisciplinary Journal of Nonlinear Science},
    volume = {34},
    number = {9},
    pages = {093113},
    year = {2024},
    month = {09},
    issn = {1054-1500},
    doi = {10.1063/5.0219961},
    url = {https://doi.org/10.1063/5.0219961},
}

@article{CZAJKOWSKI2024114994,
title = {Periodic orbit description of the blowout bifurcation and riddled basins of chaotic synchronization},
journal = {Chaos, Solitons \& Fractals},
volume = {184},
pages = {114994},
year = {2024},
issn = {0960-0779},
doi = {https://doi.org/10.1016/j.chaos.2024.114994},
url = {https://www.sciencedirect.com/science/article/pii/S0960077924005460},
author = {B.M. Czajkowski and R.L. Viana}
}

@article{SALES_10.1063/5.0222215,
    author = {Rolim Sales, Matheus and Borin, Daniel and da Costa, Diogo Ricardo and Szezech, José Danilo, Jr. and Leonel, Edson Denis},
    title = {An investigation of escape and scaling properties of a billiard system},
    journal = {Chaos: An Interdisciplinary Journal of Nonlinear Science},
    volume = {34},
    number = {11},
    pages = {113122},
    year = {2024},
    month = {11},
    issn = {1054-1500},
    doi = {10.1063/5.0222215},
    url = {https://doi.org/10.1063/5.0222215},
}

@article{HANSEN2018355,
title = {Statistical properties for an open oval billiard: An investigation of the escaping basins},
journal = {Chaos, Solitons \& Fractals},
volume = {106},
pages = {355-362},
year = {2018},
issn = {0960-0779},
doi = {https://doi.org/10.1016/j.chaos.2017.11.036},
url = {https://www.sciencedirect.com/science/article/pii/S096007791730499X},
author = {Matheus Hansen and Diogo Ricardo {da Costa} and Iberê L. Caldas and Edson D. Leonel}
}

@article{HANSEN20163634,
title = {Influence of stability islands in the recurrence of particles in a static oval billiard with holes},
journal = {Physics Letters A},
volume = {380},
number = {43},
pages = {3634-3639},
year = {2016},
issn = {0375-9601},
doi = {https://doi.org/10.1016/j.physleta.2016.09.009},
url = {https://www.sciencedirect.com/science/article/pii/S0375960116307654},
author = {Matheus Hansen and R. {Egydio de Carvalho} and Edson D. Leonel}
}

@article{BORIN2024114597,
title = {Caputo fractional standard map: Scaling invariance analyses},
journal = {Chaos, Solitons \& Fractals},
volume = {181},
pages = {114597},
year = {2024},
issn = {0960-0779},
doi = {https://doi.org/10.1016/j.chaos.2024.114597},
url = {https://www.sciencedirect.com/science/article/pii/S0960077924001486},
author = {Daniel Borin}
}

@article{OLIVEIRA20183376,
title = {Scaling invariance in a social network with limited attention and innovation},
journal = {Physics Letters A},
volume = {382},
number = {47},
pages = {3376-3380},
year = {2018},
issn = {0375-9601},
doi = {https://doi.org/10.1016/j.physleta.2018.09.034},
url = {https://www.sciencedirect.com/science/article/pii/S0375960118309940},
author = {Diego F.M. Oliveira and Kevin S. Chan and Edson D. Leonel}
}

@article{DETTMANN2012403,
title = {Escape and transport for an open bouncer: {S}tretched exponential decays},
journal = {Physica D: Nonlinear Phenomena},
volume = {241},
number = {4},
pages = {403-408},
year = {2012},
issn = {0167-2789},
doi = {https://doi.org/10.1016/j.physd.2011.10.012},
url = {https://www.sciencedirect.com/science/article/pii/S0167278911002910},
author = {Carl P. Dettmann and Edson D. Leonel}
}

@article{deFaria2016,
title = {Transport of chaotic trajectories from regions distant from or near to structures of regular motion of the {F}ermi-{U}lam model},
author = {de Faria, Nilson B. and Tavares, Daniel S. and de Paula, Wenderson C. S. and Leonel, Edson D. and Ladeira, Denis G.},
journal = {Phys. Rev. E},
volume = {94},
issue = {4},
pages = {042208},
numpages = {9},
year = {2016},
month = {Oct},
publisher = {American Physical Society},
doi = {10.1103/PhysRevE.94.042208},
url = {https://link.aps.org/doi/10.1103/PhysRevE.94.042208}
}

@article{LIVORATI2018225,
title = {Investigation of stickiness influence in the anomalous transport and diffusion for a non-dissipative {F}ermi–{U}lam model},
journal = {Communications in Nonlinear Science and Numerical Simulation},
volume = {55},
pages = {225-236},
year = {2018},
issn = {1007-5704},
doi = {https://doi.org/10.1016/j.cnsns.2017.07.010},
url = {https://www.sciencedirect.com/science/article/pii/S1007570417302605},
author = {André L.P. Livorati and Matheus S. Palmero and Gabriel Díaz-I and Carl P. Dettmann and Iberê L. Caldas and Edson D. Leonel}
}

\end{document}